\documentclass[aps,pra,twocolumn,superscriptaddress,nobibnotes,a4paper]{revtex4-1}

\usepackage[pdftex]{graphicx}
\usepackage{graphicx}
\usepackage{amsmath}
\usepackage{verbatim}
\usepackage{color}
\usepackage{bm}
\usepackage{txfonts}
\usepackage[version=4]{mhchem}
\usepackage{xcolor}
\usepackage{soul}
\usepackage{comment}

\usepackage[colorlinks=true,allcolors=blue]{hyperref} 

\begin{document}
	
\title{Hybrid integration of silicon photonic devices on lithium niobate\\for optomechanical wavelength conversion}\thanks{This work was published in \href{https://doi.org/10.1021/acs.nanolett.0c03980}{Nano Lett.\ \textbf{21}, 529--535} (2021). The source data for the figures is available at \href{https://doi.org/10.5281/zenodo.4446545}{10.5281/zenodo.4446545}.}

\author{Igor Marinkovi\'{c}}\thanks{These authors contributed equally to this work.}
\affiliation{Kavli Institute of Nanoscience, Department of Quantum Nanoscience, Delft University of Technology, 2628CJ Delft, The Netherlands}

\author{Maxwell Drimmer}\thanks{These authors contributed equally to this work.}
\affiliation{Kavli Institute of Nanoscience, Department of Quantum Nanoscience, Delft University of Technology, 2628CJ Delft, The Netherlands}

\author{Bas Hensen}
\affiliation{Kavli Institute of Nanoscience, Department of Quantum Nanoscience, Delft University of Technology, 2628CJ Delft, The Netherlands}
	
\author{Simon Gr\"oblacher}
\email{s.groeblacher@tudelft.nl}
\affiliation{Kavli Institute of Nanoscience, Department of Quantum Nanoscience, Delft University of Technology, 2628CJ Delft, The Netherlands}

\begin{abstract}
The rapid development of quantum information processors has accelerated the demand for technologies that enable quantum networking. One promising approach uses mechanical resonators as an intermediary between microwave and optical fields. Signals from a superconducting, topological, or spin qubit processor can then be converted coherently to optical states at telecom wavelengths. However, current devices built from homogeneous structures suffer from added noise and small conversion efficiency. Combining advantageous properties of different materials into a heterogeneous design should allow for superior quantum transduction devices -- so far these hybrid approaches have however been hampered by complex fabrication procedures. Here we present a novel integration method based on previous pick-and-place ideas, that can combine independently fabricated device components of different materials into a single device. The method allows for precision alignment by continuous optical monitoring during the process. Using our method, we assemble a hybrid silicon-lithium niobate device with state-of-the-art wavelength conversion characteristics.
\end{abstract}

\maketitle
	
Hybrid photonic devices have attracted significant attention for their potential in both classical and quantum information processing~\cite{Benson2011,Marpaung2019,Elshaari2020,Chu2020}. While individual materials rarely possess all desired properties, the combination of several materials allows for superior designs needed for the realization of photonic circuits that include light generation, guiding, modulation, and detection. For example, the integration of silicon photonic circuits with single photon sources~\cite{Kim2017}, two dimensional materials~\cite{Liu2011}, and classical light sources~\cite{Thomson2016} has been demonstrated, promising new capabilities beyond what is achievable with a homogeneous approach. This optimization and combination of several desired properties often comes at the expense of significantly more complex fabrication procedures, complicating the development of more advanced hybrid photonic devices. Different material systems typically react differently to chemicals or etching procedures, leading to incompatibilities in the fabrication process. Here we present a novel approach to the fabrication of hybrid devices, based on a ``pick-and-place" procedure, which is agnostic to the photonic material and compatible across a large range of different platforms. Additionally, this technique can use in-situ alignment to achieve accurate positioning without a complex imaging system. We demonstrate the capabilities of our new method by combining a silicon photonic crystal cavity with a piezoelectric lithium niobate transducer and experimentally demonstrate state-of-the-art microwave-to-optics wavelength conversion. 

Coherent conversion of quantum states between optical and microwave frequencies through a quantum transducer has become an attractive candidate to connect superconducting quantum processors. Though well-suited for local manipulation, the low frequency of superconducting circuits makes processors in distant cryostats difficult to connect~\cite{Wendin2017,Magnard2020}. A quantum transducer can solve this issue by converting quantum information into the optical domain, where it is protected against room temperature thermal noise~\cite{Lambert2020,Lauk2020}. Low-loss optical fibers can then be used to transmit information over large distances, creating a network of connected quantum processors~\cite{Kimble2008,Wehner2018}.

In particular, electro-optomechanical devices have emerged as a leading platform for realizing quantum transducers~\cite{Tian2010,Bochmann2013,Bagci2014,Balram2016,Higginbotham2018,Shao2019,Jiang2019,Forsch2020,Han2020,Jiang2020,Arnold2020,Mirhosseini2020}. In such a device, an optomechanical interaction is used to transfer quantum excitations between optical and mechanical modes~\cite{Aspelmeyer2014,Chu2020}, while a resonant electromechanical (often piezoelectric) drive can be used for efficient conversion between mechanical and microwave modes~\cite{Wu2020}. The manipulation of a quantum state of the mechanical resonator at the single phonon level has been demonstrated with both optomechanical~\cite{Chan2011,Riedinger2016,Wallucks2020} and piezoelectric interactions~\cite{Chu2017,Satzinger2018,Arrangoiz-Arriola2019}. Most recently, photons from a microwave qubit have been converted into telecom photons~\cite{Mirhosseini2020} using a hybrid Aluminum Nitride-on-Silicon-on-Insulator platform. While these first proof-of-principle experiments are highly encouraging, further improvements to the efficiency and fidelity will require even stronger piezoelectric materials.

\begin{figure}[b]
	\includegraphics[width=.98\linewidth]{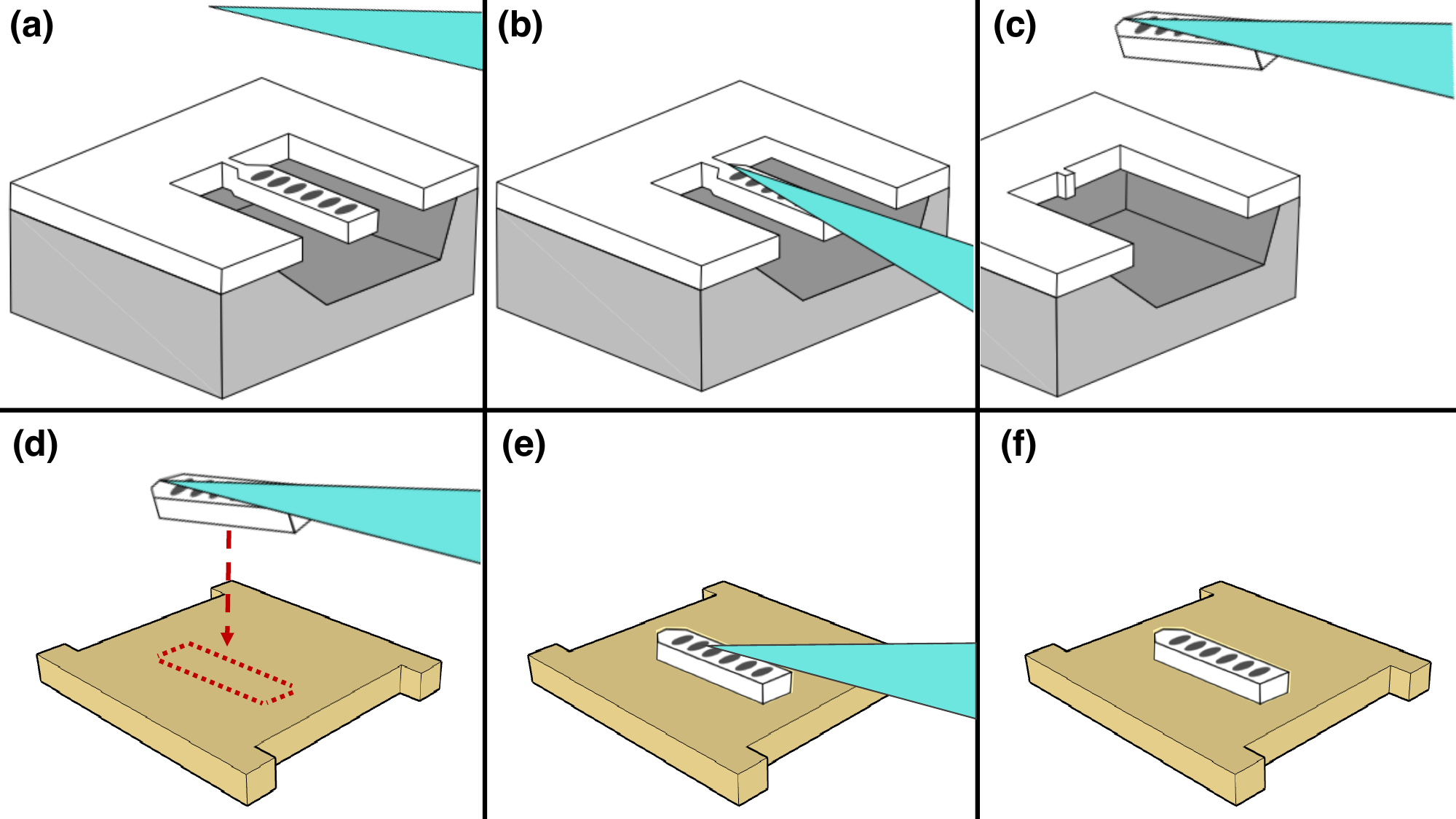}
	\caption{\textbf{Slapping:\ Pick-and-place assembly with tapered optical fibers.} A cartoon depiction of the ``slapping" procedure. (a) Photonic crystal nanobeams are patterned in a thin film of silicon (white) released from a thick oxide layer (gray). (b) A tapered optical fiber (cyan) touches the nanobeam and sticks due to van der Waals forces. (c) The thin tether of silicon connecting the nanobeam to its chip is broken through repeated motion with the fiber and the photonic device is lifted away. (d) The fiber and nanobeam are brought into close proximity with the substrate of a different chip (yellow). The optical and mechanical spectra of the nanobeam can be measured in order to map the surface and locate an optimal placement location (indicated by the red outline), in principle creating the possibility to align to various defects, such as spins, in the new material. (e) The nanobeam is then touched down onto the surface. (f) Once the device is properly assembled, the fiber is lifted.}\label{Fig1}
\end{figure}

On the one hand, state-of-the-art quantum optomechanical experiments primarily use silicon due to its high refractive index and photoelasticity, which enable large optomechanical coupling rates, as well as its small optical absorption. On the other hand, some of the most promising candidates for facilitating efficient coherent interactions between microwaves and mechanics are highly piezoelectric materials like \ce{LiNbO3} and \ce{AlN}~\cite{Shao2019,Jiang2020,Mirhosseini2020}. Therefore, combining silicon photonics with these piezoelectric materials into a single hybrid electro-optomechanical device naturally emerges as an attractive approach to establish a coherent link between microwave and optical modes~\cite{Safavi-Naeini2019}. Single-material (or homogeneous) approaches have been investigated~\cite{Jiang2020,Forsch2020}, but so far suffer from a lack of a material that fulfills the requirements for both strong piezoelectric and optomechanical coupling. To avoid this problem, optomechanical wavelength converters can use a hybrid approach~\cite{Mirhosseini2020} where a device is fabricated from multiple materials. However, this often comes at the expense of a more complex nanofabrication procedure.

In this work, we present a simple, flexible strategy for realizing such  devices. We use a pick-and-place technique~\cite{Elshaari2020} to assemble a hybrid structure from components fabricated on separate chips. A wavelength converter requires a high degree of accuracy in assembly to reach maximum efficiency. We develop a method that can achieve state-of-the-art nanoscale positioning with a straightforward procedure. Working with only a simple micropositioning stage, a microscope, and a digital camera, we utilize a tapered optical fiber to transfer a silicon photonic device onto a piezoelectric chip (cf.\ Fig.~\ref{Fig1}). Replacing commonly used tungsten tips with a tapered fiber enables one to use the photonic cavity's evanescent field as a high-precision position sensor. The simplicity and flexibility of the technique makes it well-suited for the development of novel devices, previously only possible through difficult and lengthy fabrication procedures, as well as proof-of-principle experiments with almost any combination of materials. Unlike wafer-scale bonding approaches, pick-and-place techniques enable a straightforward approach to further integration of electro-optomechanical devices with other quantum technologies without significantly changing fabrication procedures. This material-independent technique is useful for rapid prototyping and integrating new material combinations in hybrid photonic circuits. Though the surface interaction between certain materials might prevent an integration as straightforward as the one demonstrated here, it is possible to mitigate this issue by modifying the surface of the chip. Our technique is an especially attractive approach for coupling cavities to single-photon sources, as the measurement of coupling during the placement can enable an optimal positioning of the cavity. We illustrate the capabilities of our new technique by demonstrating a silicon photonic crystal cavity combined with a \ce{LiNbO3} electromechanical system, previously a difficult-to-realize material combination. We further experimentally characterize the device and demonstrate its potential for quantum transduction tasks.

\section*{D\lowercase{evice Design and Fabrication}}

\begin{figure}[t]
	\includegraphics[width=.98\linewidth]{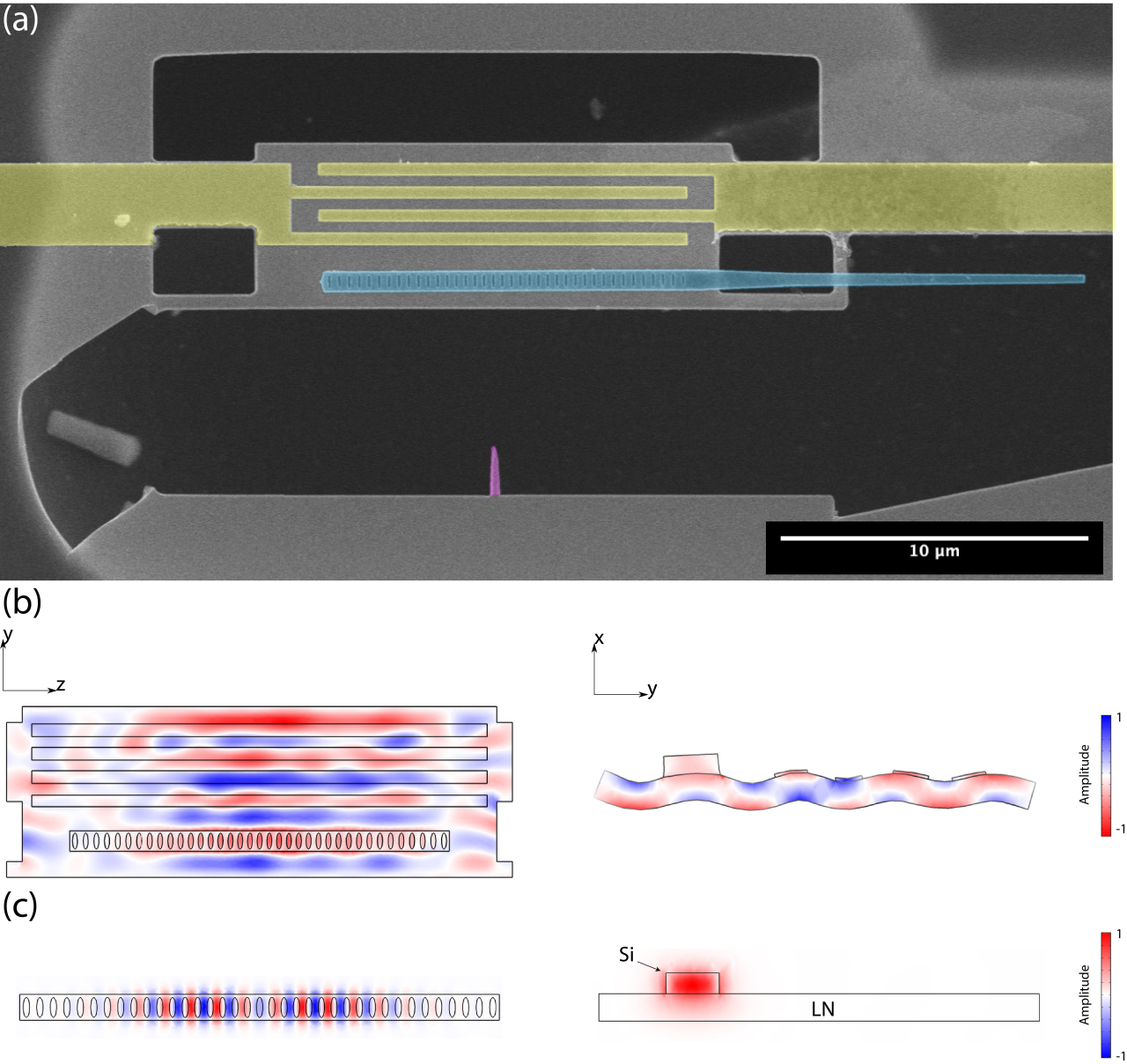}
	\caption{\textbf{Hybrid piezo-optomechanical devices.} (a) A false color SEM image of a \ce{LiNbO_3} film acoustic resonator with a silicon photonic crystal cavity placed on top. The released \ce{LiNbO_3} is light gray, the silicon nanocavity is blue, the gold electrodes are yellow, and the lithium niobate marker used for positioning is purple. The silicon cavity features a tapered waveguide that is used for coupling light from an optical fiber (not shown). (b) Simulated strain component $s_{yy}$, which is dominant for photoelastic and piezoelectric coupling. Shown is the profile of a mode of the hybrid device, top view (left) and cross-section with the mechanical deformation (right). (c) Fundamental resonant optical mode of the silicon photonic crystal cavity nanobeam, viewed from the top. The indicated axes correspond to the lithium niobate axes, while the silicon nanobeam is fabricated from a [100] wafer, with the cavity along the [110] direction.}
    \label{Fig2}
\end{figure}

Several quantum transducers using thin-film lithium niobate have shown promising results~\cite{Jiang2019,Shao2019,Jiang2020}, due to the large values of the material's piezoelectric tensor. However, these homogeneous approaches failed to achieve the high optomechanical coupling necessary for efficient transduction, as \ce{LiNbO3} has a relatively small refractive index ($n_{LN}\approx2.2$ at 1550~nm). By incorporating a silicon cavity the optomechanical coupling of a thin-film lithium niobate device can be significantly increased. In our approach, the transducer is assembled by placing a one-dimensional silicon photonic crystal nanobeam on top of a suspended lithium niobate membrane patterned with electrodes used for piezoelectric coupling (see Fig.~\ref{Fig2}a). This hybrid structure supports mechanical modes that are distributed across both materials and that couple to both optical and microwave electromagnetic fields. The membrane resonator is made from X-cut lithium niobate and is designed to support GHz-frequency Lamb wave modes along the y-axis of the crystal, as shown in Figure~\ref{Fig2}b. Through finite-element simulations, we find A0-like modes (zeroth-order asymmetric Lamb wave), where the stress inside the silicon nanobeam closely resembles the stress distribution of its breathing mode~\cite{Auld1973} (cf.\ Fig.~\ref{Fig2}c). Higher piezoelectric coupling can be expected for S0-like modes, albeit with lower mechanical quality factor due to the geometry of our device. The mechanical modes of the membrane are excited by an array of equally spaced electrodes, known as an interdigital transducer (IDT). The pitch of the electrodes determines the frequency of the excited modes, while the number of fingers sets the bandwidth~\cite{Campbell1998}. Using a large membrane with large number of IDT fingers will lead to higher coupling rates between mechanical modes and microwaves, but comes at the expense of reduced optomechanical coupling, as more effective mass is added to the mechanical mode. We design our electromechanical device with 2 finger pairs and dimensions of $0.34\times5.4\times15$~$\mu$m, with a pitch of 1.5~$\mu$m. From finite-element simulations, we expect an effective electromechanical coupling coefficient $k^2_{\textrm{eff}}\approx1\%$ for our membrane design.

The silicon nanobeam optical cavity on top of the membrane is formed by a photonic crystal mirror at each end and a tapered defect region in the middle~\cite{Chan2012}, made out of 250~nm thick silicon. Finite-element simulations of the bare silicon photonic crystal cavity nanobeam show a fundamental resonance around 1565~nm with a quality factor exceeding $2\times10^5$. When the nanobeam is placed on top of the \ce{LiNbO3} membrane its resonance is shifted to around 1595~nm (see Fig.~\ref{Fig2}c for details). We further calculate an optomechanical coupling rate of 24~kHz for an optimally positioned nanobeam.

We fabricate the lithium niobate devices using two electron-beam lithography steps. The first places electrodes on top of the membrane, and the second etches into the lithium niobate. The silicon nanobeam is fabricated in a single lithography step, with the photonic crystal cavity connected with single narrow bridge to the rest of the chip in order to be able to pick it up with the fiber during the transfer procedure (see the Supplementary Information for more details).

\section*{S\lowercase{lapping technique}}

Our approach for making hybrid photonic integrated circuits is based on the pick-and-place method, where structures are removed from a ``donor chip" made of a desired material and are transferred to a ``device chip" made of a traditional photonic material. In the past, this transfer has been accomplished using polymer stamps~\cite{Katsumi2018,Dibos2018}, atomic force microscopy (AFM) tips~\cite{Schell2011}, or tungsten nanomanipulator probes~\cite{Englund2010,Najafi2015,Mouradian2015,Zadeh2016,Kim2017,Wan2020}. In contrast, we developed a new method using an optical fiber, which we refer to as \textit{slapping}. One of the major advantages of our approach is that we can achieve high-precision positioning without sacrificing the ease of operation, because the silicon photonic devices can be directly and continuously measured during the transfer process.

Tapered optical fibers are commonly used to couple light into nanophotonic devices with high efficiency~\cite{Groeblacher2013a,Tiecke2015,Burek2017} and have also been used to rip away loosely connected photonic crystal cavities from the chip they were fabricated on~\cite{Thompson2013,Magrini2018}. Here, we extend this technique by placing the cavities onto a different substrate to create a hybrid photonic device. In our procedure (cf.\ Fig.~\ref{SFig3}), silicon photonic crystal cavity nanobeams are attached by a thin tether ($\approx 50$~nm) on a donor chip. The chip is placed on a piezo-controlled motorized stage where it is viewed from above through a $500\times$ microscope objective using a charge-coupled device (CCD) camera. A tapered fiber is placed above the donor chip at an angle of a few degrees lower than horizontal, such that the tip is nearly parallel to the surface of the donor chip. The motorized stage is then used to touch the waveguide of a photonic crystal cavity using the tapered fiber and van der Waals forces cause the fiber to adhere to the silicon. At this point, the optical and mechanical spectra of the optomechanical nanobeam can be measured with a tunable laser and photodetector using the reflected light in the optical fiber. This measurement can be used to preselect photonic structures supporting high-quality optical and mechanical modes, increasing the yield of the final devices.

Once the fiber is stuck to the nanobeam, the stages can be moved back and forth by a few microns, until the tether brakes and the nanobeam is ripped away. The donor chip is then lowered away from the fiber, and the device chip is moved into its place. The fiber is brought close to the surface of the device chip such that the silicon nanobeam and lithium niobate membrane are in focus simultaneously. A rough alignment can be done using the camera and microscope. With the evanescent field of the cavity, the environment surrounding the nanobeam can be probed, and its position with respect to the substrate can be determined. To further improve the accuracy of this alignment we include a thin marker in the \ce{LiNbO3} layer near the mechanical resonator (see Fig.~\ref{Fig2}a). By moving the nanobeam over this marker and monitoring the optical spectrum a sharp reduction in the quality factor of optical modes due to scattering in close proximity to the marker can be observed (cf.\ Fig.~\ref{SFig4}). Using this strategy, we are able to position the center of the optical cavity directly over the marker before transferring it to the mechanical resonator, which results in reliably placing our nanobeams with an accuracy of less than 100~nm, which is significantly better than what is possible with optical imaging methods. The angular alignment accuracy is typically $\ll1^{\circ}$. Additional reduction in the alignment error can easily be realized through contact sensing and monitoring of the mechanical resonance of the cavity.

Once positioned, the fiber is then lowered, until the nanobeam touches down onto the membrane. As the van der Waals force makes the nanobeam stick to the substrate more strongly than to the fiber, the fiber can be lifted away from the hybrid structure. A finalized device using this method is shown in Figure~\ref{Fig2}a. We note that this technique is not material-specific and only uses materials typically used in a laboratory capable of optical characterization.

\begin{figure}[t]
	\includegraphics[width=.96\linewidth]{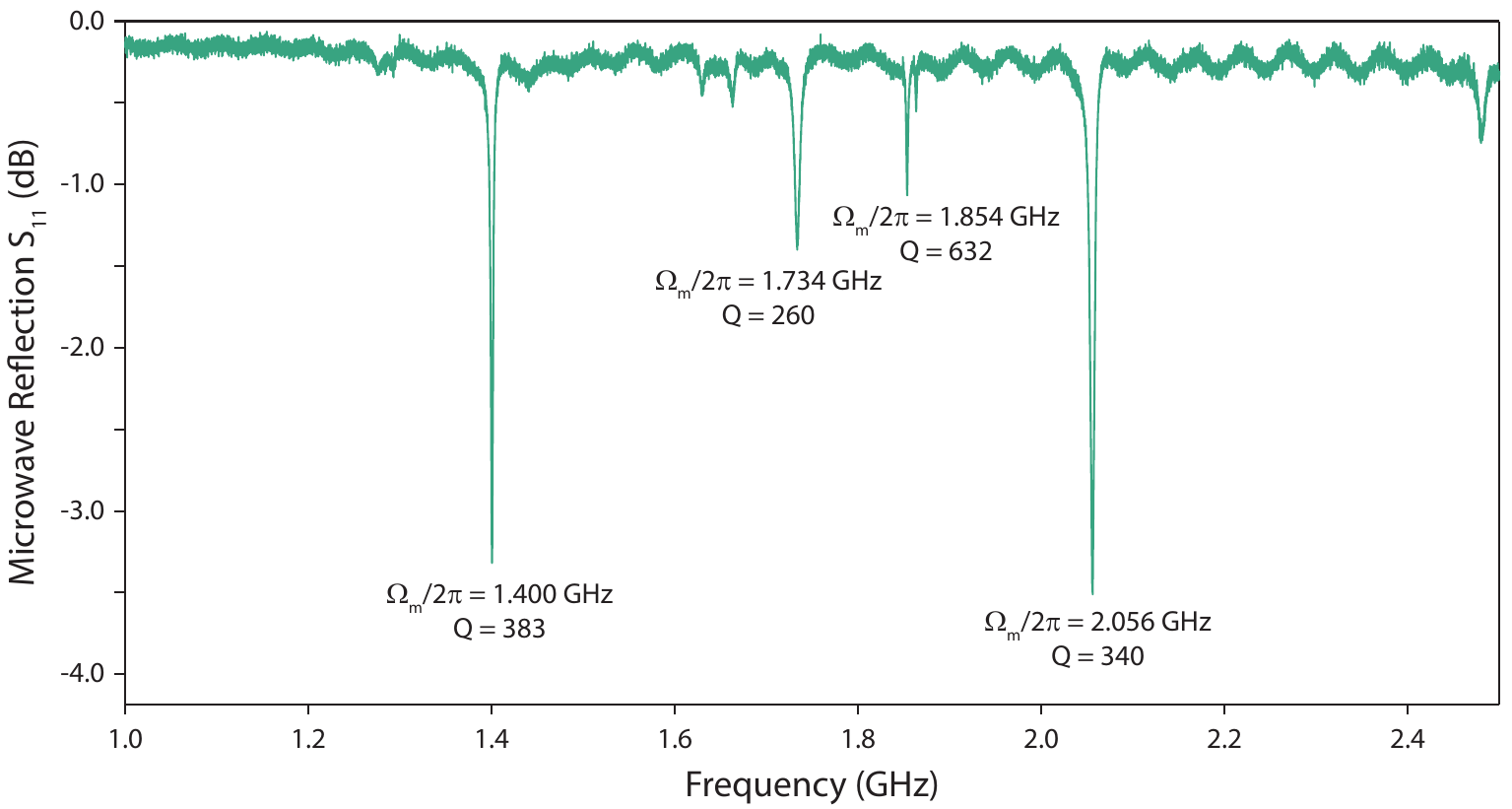}
	\caption{\textbf{Characterization of a piezo-mechanical device before slapping.}  Electromechanical reflection spectrum of a fabricated lithium niobate device before the slapping procedure. The frequency and quality factor of four significant modes are indicated. After the slapping, the mode frequencies shift slightly, and the mode line widths change (cf.\ Fig.~\ref{Fig4}).}
	\label{Fig3}
\end{figure}

\section*{R\lowercase{esults}}

After assembly, we measure the microwave and optical reflection spectra of our devices, which were all patterned with the electrodes wired to large pads in a ground-signal-ground (GSG) configuration. We measure the signal in reflection using a coaxial radio frequency (RF) GSG probe and a vector network analyzer (VNA), and the signal is normalized to the open response of the probe.

\begin{figure}[b]
	\includegraphics[width=.96\linewidth]{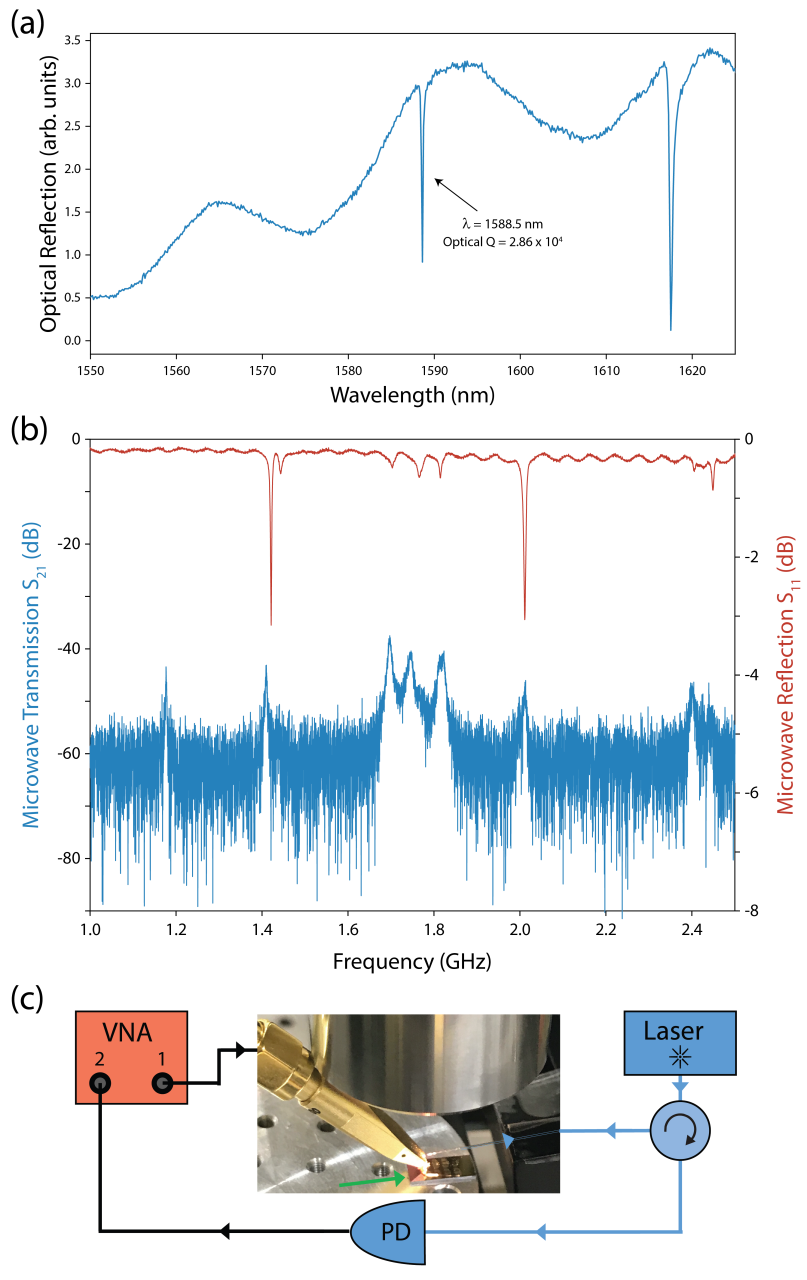}
    \caption{\textbf{Microwave-to-optical transduction using a slapped Si-on-\ce{LiNbO3} device.} (a) Optical reflection spectrum. (b) Full microwave reflection ($S_{11}$, red) and transmission ($S_{21}$, blue) signals as a function of frequency. The microwave transmission response of the device is proportional to its microwave-to-optical transduction efficiency. (c) Diagram of the experimental setup used to characterize the transduction of the device (indicated by the green arrow). PD, photodiode; VNA, vector network analyzer.}
    \label{Fig4}
\end{figure}

For our full procedure, we first characterize the lithium niobate resonator before slapping the silicon nanobeam by measuring the microwave reflection spectrum. Prominent dips in the reflection spectrum (see Fig.~\ref{Fig3}) indicate a power transfer into the mechanical modes of the device. After the slapping, the same measurement reveals that the resonances are slightly shifted, while the quality factors experience a modest decrease (Fig.~\ref{Fig4}b, red trace). We observe groups of modes spaced by approximately 330~MHz, which matches the free spectral range of the A0-like mechanical modes in our simulations. Optical spectroscopy of the hybrid device is performed using a tunable external-cavity diode laser. The reflection spectrum of the device shows two prominent resonances at 1588 and 1617~nm with a line width of 6.6 and 11.3~GHz, respectively (Fig.~\ref{Fig4}a).

We then proceed to measure microwave-to-optical transduction at room temperature. A microwave tone is swept over the device, exciting the mechanical supermodes of the hybrid nanobeam-and-membrane structure on resonance. At the same time, optical laser light detuned to the blue side of the cavity by 1.4~GHz is continuously pumped into the nanobeam, through an optical circulator and a tapered optical fiber. On resonance, the mechanical motion excited by the microwave signal will drive the optomechanical Stokes process and create a sideband resonant with the optical cavity. The optical intensity resulting from the beat note between the reflected cavity and pump photons can be measured by monitoring the light reflected from the cavity on a fast photodiode. Key parameters such as single-photon optomechanical coupling rate $g_0$ and microwave-to-optical efficiency $\eta_{eo}$ can be extracted by comparing the response of the output of the fast photodetector to the microwave input. Using a VNA yields a transmission-type $S_{21}$ measurement of microwave-to-optical transduction (Fig.~\ref{Fig4}b, blue trace).

We select the mechanical mode at 1.7~GHz as our mechanical resonance as it has the highest value in our $S_{21}$ measurement. The input optical power sent into the cavity through the tapered fiber is $I_{in}=25.3~\mu$W, while the overall received optical power at the photodiode is $I_{rec}=804$~nW. Following~\cite{Shao2019}, the single-photon optomechanical coupling rate $g_0$ can be extracted from the $S_{21}$ in the sideband resolved regime, where we use $R_{load}=50$~$\Omega$ for the impedance of the VNA ports and $R_{PD}=1300$ V/W as the responsivity of the photodiode. In our case the anti-Stokes sideband is not completely suppressed, leading to a slight under-estimation of $g_0$. The microwave-to-optical transduction efficiency $\eta_{\mu \rightarrow o}$ is the product of the single photon cooperativity $C_0$, the intracavity photon number $n_{cav}$, and the microwave and optical external coupling efficiencies ($\eta_{ext,\mu} and \eta_{ext,o}$, respectively):

\begin{equation}
	\eta_{\mu \rightarrow o}=C_0\cdot n_{cav} \cdot \eta_{ext,\mu} \cdot \eta_{ext,o}=\frac{4g^2_0}{\gamma\kappa}\cdot n_{cav}\cdot\frac{2\gamma_e}{\gamma}\cdot\frac{2\kappa_e}{\kappa}.
	\label{etafromg0}
\end{equation}

We find $g_0/2\pi \approx 10$~kHz, which matches well with the simulated value of 13~kHz for a slightly nonoptimally placed nanobeam. While the large loss rates of the optical and mechanical resonators of $6.6$~GHz and $12$~MHz, respectively, keep this device out of the $C_{o}n_{cav}>1$ regime, the relatively high external coupling efficiencies result in a photon number conversion efficiency of $\eta_{\mu\rightarrow o}\approx1 \times 10^{-7}$ at an optical power of $I_{in}=25~\mu$W in the fiber. This result is on par with homogeneous state-of-the-art \ce{LiNbO3} devices~\cite{Shao2019,Jiang2020}. On the one hand, while Ref.~\cite{Shao2019} exhibits a large microwave-to-mechanics conversion $\gamma_e/\gamma = 0.15$, it suffers from a small optomechanical coupling $g_0/2\pi=1.1$~kHz. On the other hand Ref.~\cite{Jiang2019} utilizes small mode volume nanobeam cavities to achieve $g_0/2\pi \approx 80$~kHz but at the expense of low $\gamma_e/\gamma=0.001$. Our platform combines the strengths of both, with $g_0/2\pi=10$~kHz and $\gamma_e/\gamma=0.015$. Compared to a recent AlN on Si design~\cite{Mirhosseini2020}, our device also exhibits a significantly higher piezoelectric coupling. Following~\cite{Jiang2019}, we estimate that the coupling of our device to a microwave LC resonator of the same impedance would yield about an order of magnitude higher coupling $\sim$2$\pi\times 40$~MHz. An additional advantage of our design is the better thermal anchoring, resulting in a reduced susceptibility to optical absorption induced heating.

We have demonstrated a new approach for wavelength conversion that is on par with other recent state-of-the-art devices. We have done this by successfully integrating Silicon and LN into a single hybrid electro-optomechanical device using a new and simplified variation of the pick-and-place technique. Our work shows that placing can be done on a suspended membrane, while forming good mechanical contact. This confirms that this simple technique is relevant for fabricating optomechanical devices. As the material combination used in this work is superior for the optomechanical quantum transduction task than any other approach to date, additional adjustments to our platform can be used to significantly improve the performance of this new class of devices by several orders of magnitude. Coupling between microwave and mechanical modes can be increased by using an on-chip impedance matching circuit and switching to S0-like modes. In our \ce{LiNbO3} membranes these modes suffered from small quality factors, which can be overcome by adjustments to the fabrication procedure, as similar designs have demonstrated S0-like modes with much larger $Q$ factors~\cite{Gong2013}. This will enable us to use a smaller membrane (while preserving the piezo $k^2_{eff}$), which will increase the optomechanical coupling rate. Further improvements in the fabrication will lead to higher optical quality factors and, hence, to a better overall efficiency, which has already been demonstrated in a similar structure~\cite{Witmer2016}. Our new slapping approach is also directly applicable to the precision positioning of photonic crystal cavities to spins in a substrate, such as for color centers in diamond~\cite{Hausmann2013} and rare-earth ions~\cite{Kindem2020,Raha2020}. While a deterministic pick and place between nitrogen-vacancy centers and photonic crystal cavities has been demonstrated~\cite{Englund2010}, our approach using a tapered fiber for simultaneous high-efficiency readout and pick-and-place procedure with in situ alignment simplifies the assembly process significantly.

\begin{acknowledgments}
We thank Robert Stockill, Moritz Forsch, and Roald van der Kolk for valuable discussions. This work is supported by the Foundation for Fundamental Research on Matter (FOM) Projectruimte Grant (16PR1054), the European Research Council (ERC StG Strong-Q, 676842), and by the Netherlands Organization for Scientific Research (NWO/OCW), as part of the Frontiers of Nanoscience program, as well as through Vidi (680-47-541/994) and Vrij Programma (680-92-18-04) grants. B.H.\ acknowledges funding from the European Union under a Marie Sk\l{}odowska-Curie COFUND fellowship.
\end{acknowledgments}

\setcounter{figure}{0}
\renewcommand{\thefigure}{S\arabic{figure}}
\setcounter{equation}{0}
\renewcommand{\theequation}{S\arabic{equation}}

\pagebreak
\clearpage

\section*{Supplementary Information}

\subsection{Fabrication}

\begin{figure}[b]
	\includegraphics[width=.98\linewidth]{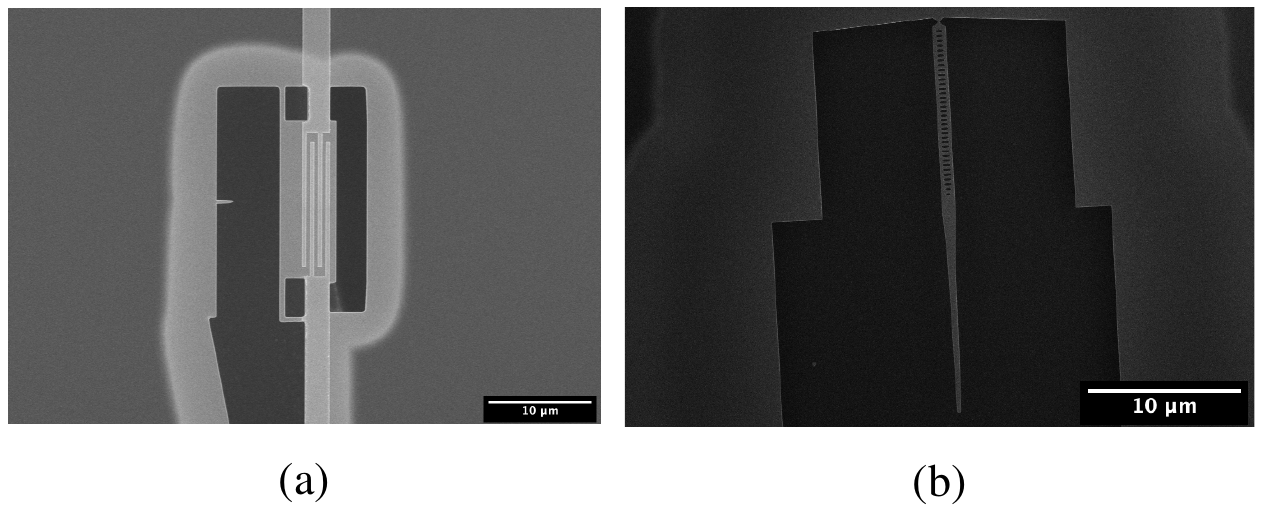}
    \caption{\textbf{Fabrication.} (a) A SEM image of a lithium niobate resonator on the device chip before the slapping procedure. (b) A SEM image of a silicon photonic crystal cavity nanobeam on the donor chip before slapping. The structure is connected to the chip by the tether at the top.}
    \label{SFig1}
\end{figure}

We begin with a Lithium Niobate on Silicon (LNOS) $5 \times 10$~mm chip, consisting of a 340~nm layer of X-cut, single crystal lithium niobate doped with 5~mol$\%$ \ce{MgO} bonded to a 500~$\mu$m thick silicon substrate. Electrodes are defined by first patterning a 280~nm thick coating of electron beam resist and then evaporating an adhesion layer consisting 5~nm of chromium and 15~nm of platinum under a 40~nm layer of gold. The evaporation step is followed by lift-off, where the chip is placed in a beaker of anisole heated to $80~^{\circ}$C and set in an ultrasonic bath for five minutes. Next, the second electron beam lithography step defines the shape of the thin film resonators in a thick 550~nm coating of CSAR62.13 resist. The exposed lithium niobate is etched through by a large cross-sectional beam of argon ions incident at a 90-degree angle to the surface of the chip. This process etches through the lithium niobate membrane and results in a sidewall angle of $75^{\circ}$. The devices are released using a dry \ce{SF6} reactive ion etch which isotropically etches the silicon through the holes in the lithium niobate film. The resist is not stripped until after the underetching step in order to prevent contamination of the etcher. At this point, resist removal is achieved by ashing in oxygen plasma. Finally, we employ an inorganic cleaning step to remove lithium niobate redeposited during the physical etching. We prepare a mixture similar to RCA-1, a 2:2:1 ratio of \ce{NH4OH} [$28\%$] : \ce{H2O2} [$31\%$] : \ce{H2O}. The mixture is heated to $85~^{\circ}$C and mixed with a stir bar for 5 minutes before our samples are placed in for another 5 minutes. A fabricated device is shown in Fig.~\ref{SFig1}a. The silicon nanobeam fabrication is described in detail in~\cite{IgorPhD}, with the liquid HF undercut being replaced with vapor HF. A fabricated nanobeam device is shown in Fig.~\ref{SFig1}b.

\subsection{Redeposition of Lithium Niobate}

\begin{figure}[b]
	\includegraphics[width=.98\linewidth]{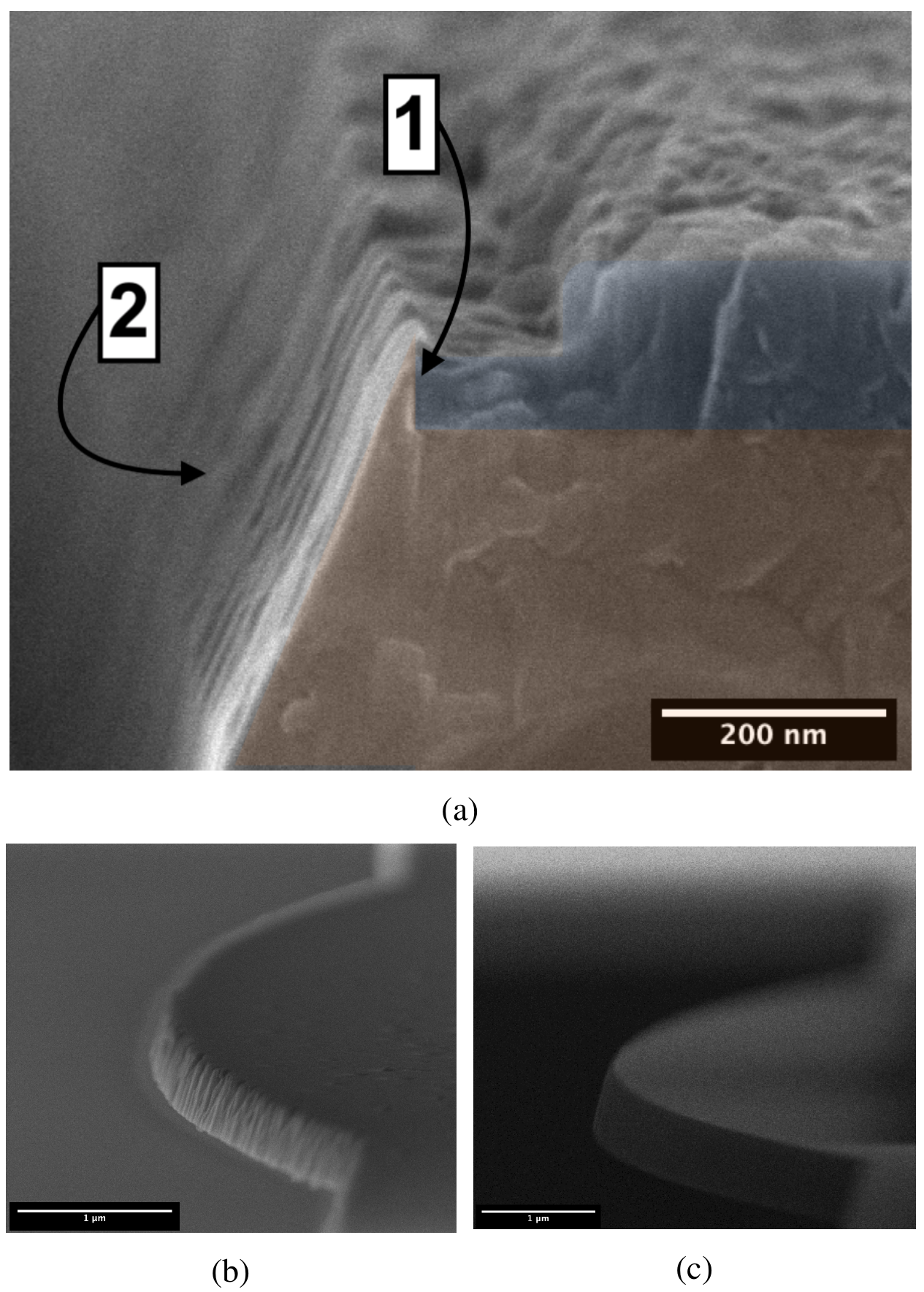}
    \caption{\textbf{Redeposition of Lithium Niobate.} (a) A SEM image of lithium niobate after ion milling, with the profile in false color. \ce{LiNbO3} is orange and the remaining resist is blue. After milling through the \ce{LiNbO3}, angular features about 45~nm wide and 60~nm tall are redeposited above the \ce{LiNbO3} layer of our sample, indicated by Marker 1. Marker 2 points to the resulting rough sidewalls. (b) A curved structure in a suspended thin film of \ce{LiNbO3} shows the same issues. (c) The same device after RCA-1 cleaning. The sidewalls appear smooth and no redeposited features are present.}
    \label{SFig2}
\end{figure}

Despite being a widely used material across multiple industries for decades, the nanofabrication of lithium niobate is considered difficult. There is no known chemical etch for lithium niobate that produces desirable devices with acceptable sidewalls and low surface roughness without unwanted contamination. This leaves as the only viable alternative physical etching. Inspired by Refs.~\cite{Wang2013,Wang2014,Arrangoiz-Arriola2018}, we process our lithium niobate using argon ion beam etching (also called ion milling) in an inductively coupled plasma etcher. However, ion milling lithium niobate causes amorphous \ce{LiNbO3} redeposition, creating unwanted features and increasing sidewall roughness as seen in Figs.~\ref{SFig2}a and~\ref{SFig2}b. We mitigate this problem first by optimizing the plasma etching parameters to minimize redeposition and then by chemically cleaning our sample. The results are shown in Fig.~\ref{SFig2}c. The chemical cleaning step is based on RCA-1, which is typically used on silicon devices~\cite{Itano1993} and found to also be effective on lithium niobate~\cite{Cai2019}.

\begin{figure*}[t]
	\includegraphics[width=.98\linewidth]{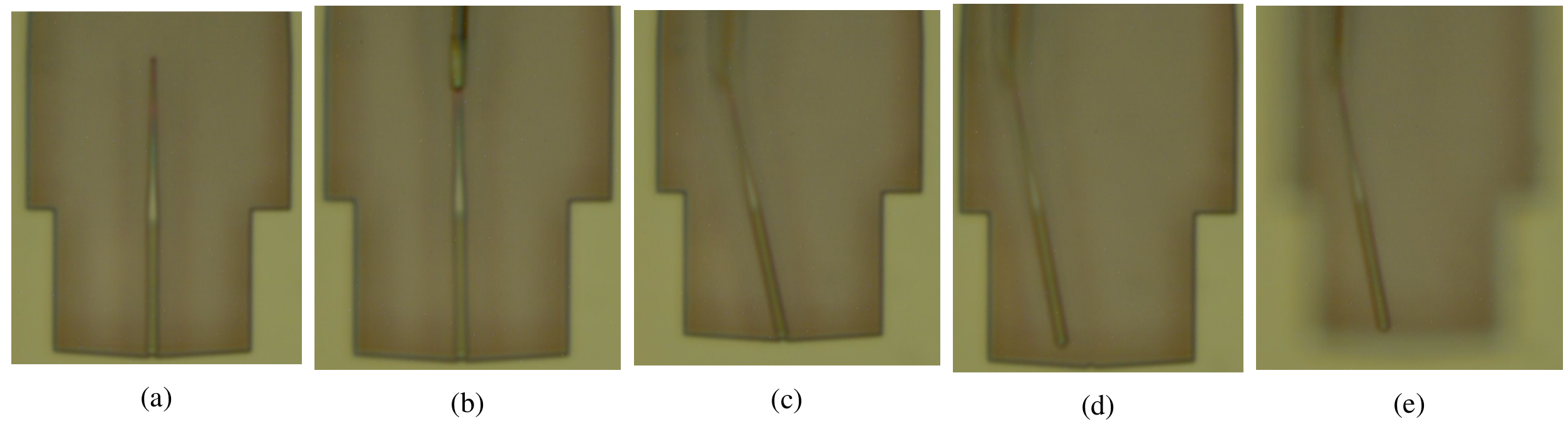}
	\caption{\textbf{Slapping Pick-up.} (a) An optical microscope image of a photonic crystal cavity nanobeam before slapping. (b) A tapered optical fiber is placed in contact with the tapered waveguide of the nanobeam, allowing to measure its resonance with high efficiency. (c) The fiber is then moved around, dragging the nanobeam with it until the tether is broken. (d) With the tether broken, the nanobeam can be moved away from the chip. (e) Finally, the fiber is lifted from the donor chip. The nanobeam is now ready to be slapped onto a new device.}
    \label{SFig3}
\end{figure*}

\subsection{Slapping Procedure}

The slapping technique consists of first picking up and then slapping down a structure from a donor chip onto a device chip. The pick-up procedure begins by placing a silicon donor chip under a microscope and locating a photonic crystal cavity nanobeam (Fig.~\ref{SFig3}a). In this image, the device is connected to the chip by a thin tether at the bottom (see SEM in Fig.~\ref{SFig1}b). Next, a tapered optical fiber is lowered until the device and the tip of the fiber are simultaneously in focus. The tapered fiber is then placed on the tapered waveguide of the nanobeam (Fig.~\ref{SFig3}b). It is easy to distinguish when the fiber is touching the nanobeam, as the fiber changes color on contact. At this stage, the optical and mechanical spectra of the nanobeam device can be measured in order to confirm that it is functioning as intended. When the fiber is moved to the left or the right, the nanobeam adheres and pivots around the tether (Fig.~\ref{SFig3}c). We find that sweeping the fiber left and right by a few microns reliably breaks a 50-100~nm thick tether. Once the tether is broken, the nanobeam can be pulled away from the donor chip (Fig.~\ref{SFig3}d) and finally be lifted away (Fig.~\ref{SFig3}e).

Next, after picking up the nanobeam, it can be placed back down wherever desired. Slapping the nanobeam only requires that the structure sticks better to the new substrate than it does to the tapered fiber tip. This is usually true for any material, as the nanobeam is picked up in such a way that it contacts only a small fraction of the nanobeam's surface area. When the nanobeam is slapped down, its entire bottom surface touches the substrate and thus it adheres. We found that once slapped, the structures' position can be further adjusted by pushing them across the substrate with the tapered fiber after slapping.

\subsection{Increasing Slapping Accuracy with In-Situ Optical Sensing}

The accuracy of slapping an arbitrary structure is in principle limited by the optical diffraction limit and practically by operator error. However, in our approach we are picking up optical cavities with an optical fiber. This allows us to significantly improve the accurate placement by measuring the photonic crystal cavity during the slapping procedure.

\begin{figure}[b]
  \centering
  \includegraphics[width=.98\linewidth]{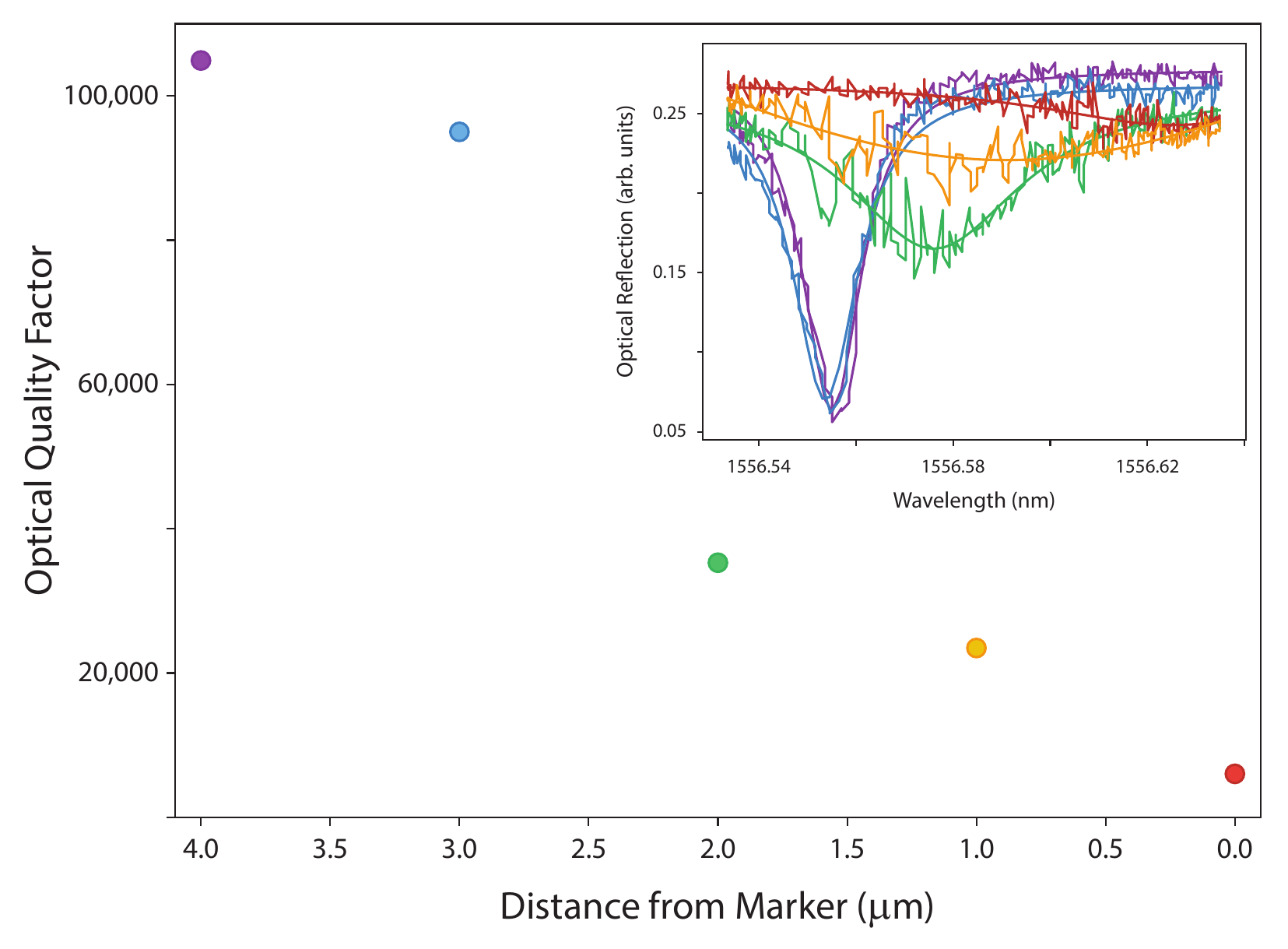}
  \caption{\textbf{In-Situ Sensing.} Measurement of the silicon optomechanical crystal nanobeam attached to an optical fiber enables increased positioning accuracy with respect to an on-chip marker. The quality factor of the fundamental optical mode is significantly reduced as the nanobeam approaches the marker. Inset:\ Corresponding optical reflection spectrum of the nanobeam at each point.}
  \label{SFig4}
\end{figure}

In order to demonstrate and characterize the in-situ alignment procedure of the position of the cavity with respect to a feature on a device chip, we monitor the resonance as we approach a silicon marker. The marker is simply a 270-nm thick silicon cantilever, similar to the lithium niobate marker visible in the bottom of Fig.~\ref{Fig2}b. In order to sense the marker, the nanobeam is placed a few hundred nanometers above the device chip while the optical resonance is measured on reflection using the tapered fiber. Scattering drastically reduces the quality factor of the optical cavity when the marker is near the defect region of the photonic crystal. In addition, the resonances are slightly redshifted close to the marker (see Fig.~\ref{SFig4}). Both effects are clear, as the redshift and the linewidth of the fundamental optical mode increase with proximity to the marker. By fitting the data to find the position of maximum absorption, we can locate the center of the photonic crystal cavity directly over the on-chip marker. Using this technique, we are able to position a nanobeam with an accuracy of better than 100~nm, well under the optical diffraction limit (100s of nanometers in this case).

\subsection{Photonic Crystal Design}

The cavity is formed by two photonic crystal mirrors at each end and a tapered region in the middle, designed and fabricated according to~\cite{Witmer2016}. The unit cell periodicity of the mirror is $a_m = 338$~nm and hole axes are $r_1 = 78$~nm and $r_2 = 258$~nm, while the width of the nanobeam is $w = 650$~nm and the thickness of the silicon layer is $250$~nm. The lattice spacing is quadratically tapered to a value $a_d = 0.91$~nm and $a = 307.6$~nm in the central defect region. Through the simulation of the optomechanical coupling rate $g_0$ of various mechanical modes to the photonic cavity, we confirm that the coupling inside the lithium niobate is small and hence can be neglected, as most of the optical mode is contained inside of the silicon. The relatively low quality factor of the fabricated cavities can be attributed to poor etching of the silicon.

\subsection{$g_0$ of S0 mode}

\begin{figure}[h]
  \includegraphics[width=.98\linewidth]{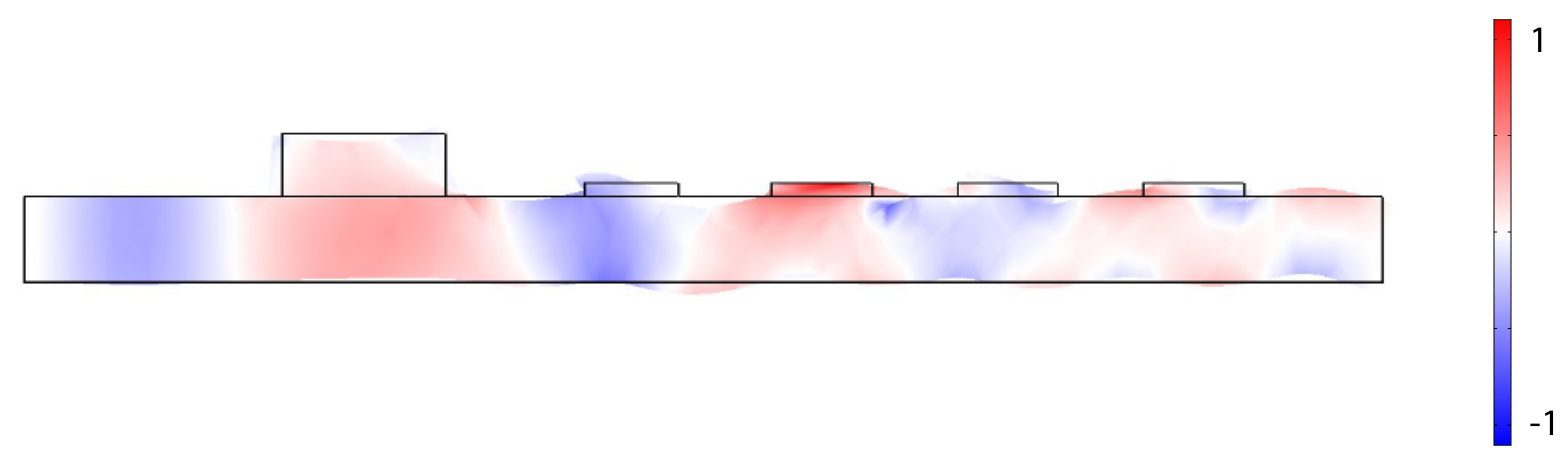}
    \caption{\textbf{S0-like mode of the device}. This mode gives the best performance, as most of the device is either coupled to optics or microwaves through high photoelastic/piezoelectric coefficients.}
    \label{SFig6}
\end{figure}

We calculate the optomechanical coupling rate for S0-like modes of our hybrid device (Fig~\ref{SFig6}), as these modes have higher piezoelectric coupling. We obtain $25$~kHz for a $3.68$~GHz mode.

\subsection{Estimation of $g_0$}

Here we like to sketch how we estimate $g_0$ from our experimental data. Light in the photonic cavity can be decomposed into three fields:\ the pump and the two optical sidebands shifted from the pump by the RF drive frequency. Expressions for the amplitudes of these waves ($A_0$, $A_+$ and $A_-$) are derived in~\cite{Shao2019}:

$$A_0=\frac{\sqrt{\kappa_e} A_{in}}{i\Delta+\kappa/2}$$
$$A_-=\frac{-iG A_{0}}{i(\Delta+\Omega_d)+\kappa/2}$$
$$A_+=\frac{-iG A_{0}}{i(\Delta-\Omega_d)+\kappa/2}$$

\noindent where $\Delta=\omega_c-\omega_L$ is the detuning of the laser with respect to the cavity. $G=2g_0\sqrt{\frac{P_{in}}{\hbar\Omega_m}}\sqrt{\gamma_e}/\gamma$, with $\Omega_d$ and $P_{in}$ being the frequency and power of the microwave drive of the mechanical resonator, respectively, and $\lvert A_{in} \lvert^2=\frac{I_{in}}{\hbar\omega_L}$. Light coming out of the cavity is then given by:

$$A=A_{in}-\sqrt{\kappa_e}A_0-\sqrt{\kappa_e}A_-e^{i\Omega_d t}-\sqrt{\kappa_e}A_+e^{-i\Omega_d t}$$

After propagating to the detector with an efficiency $\eta=0.28$, we detect the signal at the mechanical frequency with an amplitude $U=R_{PD}\hbar\omega_c\lvert A\lvert ^2$. We then calculate the power that the VNA receives as $P_{out}=U^2/(2R_{load})$, and finally obtain $S_{21}=P_{out}/P_{in}$. Using this relationship between $g_0$ and $S_{21}$ we can then numerically calculate $g_0$.

\subsection{$V_{\pi}$ of a Hybrid Device}

\begin{figure}[b]
  \includegraphics[width=.9\linewidth]{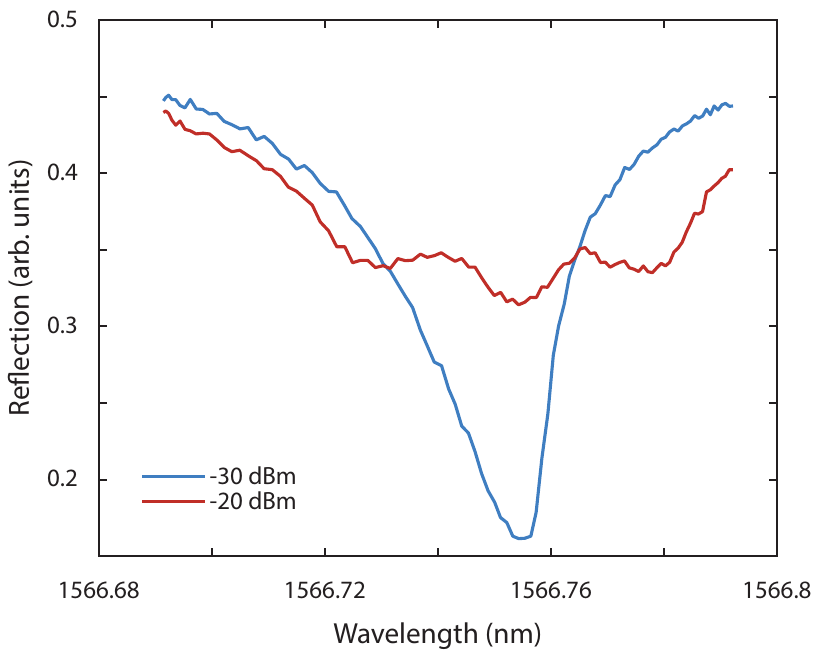}
    \caption{\textbf{$V_{\pi}$ of a slapped hybrid device}. Blue trace is taken with an RF power sent to IDTs of -30~dBm at 3.06~GHz, while the red trace corresponds to -20~dBm.}
    \label{SFig7}
\end{figure}

We have characterize another device with higher mechanical frequency and slightly smaller optical linewidth. This enables us to clearly see the modification of optical reflection spectrum due the presence of RF drive on the IDTs (see Fig.~\ref{SFig7}), as done in Ref.~\cite{Shao2019}. This makes the resonance frequency of the cavity to oscillate with respect to the laser at the drive frequency. In the cavity's frame of reference the laser frequency is modulated, giving rise to sidebands at the RF drive frequency. We use the formula given in~\cite{Shao2019}
\begin{equation}
S_{21}=\left(\frac{\pi R_{PD} I_{rec}}{V_{\pi}}\right)^2
\end{equation}
to estimate the $V_{\pi}\approx 50$~mV of our device.
	
\end{document}